\UseRawInputEncoding
\documentclass[
bibnotes,
 amsmath,amssymb,
 aip,
]{revtex4-2}
 \usepackage{graphicx}
  \usepackage{enumerate}
\usepackage{url}
\newcommand{\bey}{\begin{eqnarray}}
\newcommand{\eey}{\end{eqnarray}}

\begin{document}
\title{Gravitational Decoherence: A Thematic Overview}

\author{Charis Anastopoulos\footnote{anastop@upatras.gr}\\
{\small Department of Physics, University of Patras, 26500 Patras, Greece}\\
{\small and} \\
Bei-Lok Hu\footnote{blhu@umd.edu}\\
{\small Maryland Center for Fundamental Physics and Joint Quantum Institute,} \\
{\small University of Maryland, College Park, Maryland 20742-4111, USA}
}

\begin{abstract}
 Gravitational decoherence (GD) refers to the effects of gravity in actuating the classical appearance of  a quantum system. Because the underlying processes involve issues in general relativity (GR), quantum field theory (QFT) and quantum information, GD  has fundamental theoretical significance. There is a great variety of GD models, many of them involving physics that diverge from GR and/or QFT.  This overview has two specific goals along one central theme: (i) present theories of GD based on GR and QFT and explore their experimental predictions;  (ii)  place other theories of GD under the scrutiny of GR and QFT, and point out their theoretical differences.  We also describe how GD experiments in space in the coming decades can provide evidences at two levels: a) discriminate alternative quantum theories and non-GR theories;  b) discern whether gravity is a fundamental or an effective theory.\\

\noindent --- {\it Invited paper in a Special Issue of AVS-QS in celebration of Sir Roger Penrose's 2020 Physics Nobel Prize Award}.

\end{abstract}

\date{November 3, 2021}

\maketitle


\section{Introduction}

Gravitational decoherence (GD) is a fundamental issue of theoretical physics because gravitation is a universal interaction and decoherence is an essential factor in how a quantum  system assumes some classical  behavior.  GD encompasses the issues of both the quantum-to-classical  and the microscopic-to-macroscopic transitions. The quantum-to-classical transition is the description of a quantum system in terms of classical physics. The microscopic-to-macroscopic transition is the accurate description of systems with $10^{23}$ degrees of freedom in terms of a few observables, as in thermodynamics.

By their very nature, GD and gravitational entanglement (GE) also involve basic quantum information issues, starting with the effects of gravity on  the quantum-to-classical transition \cite{Karol, Diosi0, Penrose}, and continuing to the  construction and probing of gravitational cat states \cite{AnHu15}. The possibility of direct experimental feedback about a non-trivial interplay between gravity and quantum theory is what makes those phenomena so interesting  from a theoretical viewpoint.

Note that one does not need to appeal to theories of quantum gravity at the Planck scale to see the contradictions between the quantum and gravitation,  to ask  questions about the quantum nature  of gravity and how the quantum-informational features show up.  Their manifestations are already present in low energy physics, namely, in the nonrelativistic (NR) and weak field (WF) limits of general relativity (GR) and quantum field theory (QFT).  This is the regime we live in where laboratory experiments on Earth and in space  can offer invaluable observational data to cross-examine the above-mentioned theoretical issues of fundamental importance.

\subsection{Quantum Decoherence}

For a long time since the advent of quantum mechanics, the issues  of the quantum-to-classical transition and of the  emergence of the classical world  have been a subject of curiosity for debates among the philosophers of science. Physicists reclaimed the subject in the 80s with full rigor in three lines of development.

\medskip

{\bf A)  Consistent/ Decoherent histories.} This program was developed by Griffiths \cite{Griffiths}, Omn\'es \cite{Omnes} and Gell-Mann \&  Hartle \cite{GeHa}. The key idea is that quantum theory must be expressed in terms of histories, i.e., sequences of properties of a system at different moments of time, rather than in terms of evolving single-time quantum states. The decoherent histories (DH)  program places great emphasis on the logical structure of propositions about histories, which leads to a rigorous implementation of the crucial notion of coarse-graining. As a result, the DH program offers the  most sophisticated account of the quantum-to-classical transition in a unified way  with the microscopic-to-macroscopic transition. However, DH guarantees neither the uniqueness nor the stability of the classical world of our experience \cite{KeDo}, while it offers only minor improvements over Copenhagen quantum mechanics on the measurement problem \cite{DHmeas}.

\medskip

 {\bf B) Environment-Induced Decoherence} (EID). This program originates from Zurek \cite{Zurek},  and Joos and Zeh \cite{JoZe}, following an earlier work by Zeh \cite{Zeh}. The key idea is that quantum systems classicalize as a result of their interaction with the environment. This program went hand in hand with the development of the theory of open quantum systems. Most prominent are the master equations that built on the work of Schwinger \cite{Schwinger} and Feynman and Vernon \cite{FeVe} in the 1960s  for the ubiquitous quantum Brownian motion, to the  Caldeira-Leggett Markovian master equation \cite{CaLe}  of the 1980s which is valid  for Ohmic environments at high temperatures, and to the Hu-Paz-Zhang  non-Markovian master equation of the 1990s \cite{HPZ} that is valid for a general environment at all temperatures. The applicability of the program is limited to setups that admit a natural system-environment split.   And, despite its many successes in matching with experimental outcomes,
  it does not provide, by itself, a  satisfactory solution to the quantum measurement problem \cite{decohmeas}.

\medskip

{\bf C) Wavefunction Collapse models}.  This program introduces a scale that separates the microscopic from the macroscopic and postulates that  quantum mechanics as we know it which applies to a plethora of natural phenomena from subatomic scales up, no longer holds for macroscopic phenomena. There are two major directions. The first direction originates from Ghirardi, Rimini and Weber \cite{GRW} and Pearle \cite{Pearle} (hence, GRWP), and it culminates in the formulation of dynamical reduction or continuous state localization (CSL) models \cite{various}. The second direction, exemplified by Diosi \cite{Diosi0} and Penrose \cite{Penrose} (DP model), focuses on wave-function collapse due to gravity, and we will discuss it separately.

Unlike the previous two, this approach postulates a fundamental modification of quantum theory, and for this reason we will refer to the theories in this class as alternative quantum theories (AQT).   CSL models start from the pragmatic aim of introducing  a  scale  where macroscopic phenomena overtake the microscopic quantum mechanical behavior through a mathematically consistent modification of quantum dynamics. This mostly works in non-relativistic physics.  Still, the whole procedure is {\em ad hoc}, with no fundamental explanation, and one encounters grave problems when attempting to formulate a relativistic version of CSL models appropriate for QFT. Despite significant progress in recent years \cite{CSLQFT}, a  CSL theory that is fully consistent with special relativity has yet to appear \cite{JGB}.

\subsection{Gravitational Decoherence}
The main body of work on gravitational decoherence began a decade later. We shall mention the main lines of development in the next section but give an overall perspective here.

One special feature in this research subfield is a widened scope of possibilities, invoking novel features from gravity as well as  quantum theory.  We observe that both the DH and the EID programs do not call on alteration of GR or QM.  DH allows for generalizations of quantum theory for systems that are difficult to treat with usual methods, or for the definition of more elaborate observables, but the physical predictions in the domain of current theories are the same with the standard ones. EID  highlights the role played by the environment in determining the classical outcomes of a quantum system. As such, it requires tools and concepts in open quantum systems and non-equilibrium statistical mechanics, but it does not need to alter  existing theories at the fundamental level.

With gravitational decoherence, we see proposals which call for the alteration of quantum mechanics, such as in the GRWP and DP theories.  The motivation ranges from pragmatic considerations as in CSL to  the `gravitisation' of quantum mechanics, proposed by Penrose \cite{Penrose2}.   Now that gravity enters the picture many proponents find it convenient to use GR, even altering it,  to explain the quantum-to-classical transition. This is done  by invoking fluctuations either in the manifold structure or in the spacetime metric,  under the labels of `intrinsic', `fundamental', or `quantum gravity' decoherence. We shall point out how such models infringe on  GR in subtle ways.

\subsection{Features of this overview}

An excellent review is available on this subject \cite{gravdec} which covers a broad range of topics with prominent representation of CSL and AQTs. For this reason, we shall narrow our goals in this overview to two: (i) Present theories of gravitational decoherence based on GR and QFT  and their experimental predictions. (ii)  Place other theories of GD under the scrutiny of GR and QFT, and point out the theoretical differences.  In fact, we would urge all proponents of new and old alternative theories to present,  alongside with what they consider as their proposals' attractive features, also a self-scrutiny against  GR and QFT  as benchmarks. The reason is simply that, these two theories are the pillars of modern physics, and they serve as the yardstick by which to distinguish between proposals based on "known physics" or conservative extensions thereof  and proposals based on "new physics" which violate either GR or QFT or both, in ways big or small.

In Sec. 2, we present the main theories of gravitational decoherence classified according to the physical origins of decoherence. Decoherence may originate from the fundamentally classical nature of gravity, from quantum gravity processes or from spacetime fluctuations. In Sec. 3, we present the Anastopoulos-Blencowe-Hu (ABH) theory of gravitational decoherence \cite{AHGravDec, Miles} that fully lies within the scope of GR and QFT. In Sec. 4, we briefly review estimates for gravitational decoherence effects, mainly comparing the predictions of the DP and the ABH model.

\section{Theories of gravitational decoherence}

\subsection{Decoherence Vs. dephasing}

The term ``gravitational decoherence" is sometimes used in the literature when the authors actually refer to dephasing due to gravity \cite{dephasinggravity}.
Decoherence is an irreversible process, while dephasing may also occur without irreversibility.    A classic case of mere dephasing without decoherence is provided by the spin echo experiments \cite{hahn}. In these experiments, phase information is apparently lost, but a simple operation on the system will recover this information. In absence of dissipation (an irreversible process), the recovery of information is complete \cite{AnSav11}.  In decoherence,  the phase information is lost forever, or at least for a very long  time (of the order of the Poincar\'e recurrence time of the environment). Another analogy in terms of energy is the following. Mere dephasing  is in the nature of  Landau `damping'  a la the Vlasov equation which is of the Hamiltonian type,  decoherence is in the nature of the nonunitary Boltzmann equation with real physical dissipation.

Here, we will only be concerned with models of actual irreversible decoherence.

 \subsection{ Decoherence from classical gravity}

 This is the oldest approach and most influential approach to gravitational decoherence. It originates with Karolyhazy \cite{Karol} in the 1960s. An underlying idea is that gravity is a fundamentally classical channel of interaction. This implies that gravity can act as an agent of decoherence in quantum system.

 \subsubsection{The Diosi-Penrose theory}

According to Penrose \cite{Penrose}, decoherence is a plausible consequence of  the fundamental incompatibility between GR and quantum theory, especially in their respective treatments of time. Time in quantum theory is essentially classical and forms a background to all quantum phenomena, while in GR time is fundamentally dynamical as it is determined by the spacetime metric that is a dynamical observable.

  The contradiction between the role of time in GR and in quantum theory is most emphatically manifested  when considering superposition of macroscopically distinct states for matter. By Einstein's equations, each component of the superposition generates a different spacetime. Since there is no canonical way of relating time parameters in different spacetime manifolds, there is a fundamental ambiguity in  the time parameter of the evolved quantum state. According to Penrose, this ambiguity is manifested even at low energies, when the gravitational interaction can be effectively described by the Newtonian theory. It may be expressed as a  mechanism of gravity-induced decoherence for superpositions of  states with different mass densities $\mu_1(x)$ and $\mu_2(x)$. Penrose proposed that the time-scale of decoherence is of the order of   the gravitational self-energy of the difference in the two mass densities.

Diosi \cite{Diosi0} constructed a master equation in which he postulated a collapse term, with noise correlator proportional to gravitational potential. The master equation is
 \begin{eqnarray}
 \frac{\partial \hat{\rho}}{\partial t} = -i [\hat{H}, \hat{\rho}] - \frac{G}{2} \int d{\pmb r} d {\pmb r'} \frac{[\hat{\mu}({\pmb r}),[\hat{\mu}({\pmb r}'),\hat{\rho}]]}{|{\pmb r} - {\pmb r}'|}, \label{DPME}
 \end{eqnarray}
 where $\hat{\mu}({\pmb r})$ is the mass density operator.
 Key value in this master equation is  that predictions do not involve any free parameters, except for a high-frequency cut-off $\Lambda$ that is necessary for the definition of the mass density operator.  The natural  scale of $\Lambda$ for non-relativistic particles is the nuclear scale, in the sense that $\Lambda^{-1}$ corresponds to a nucleon's radius, nuclei being the simplest particles most affected by this decoherence mechanics. The decoherence rate is typically of the order of $\frac{1}{\Delta E}$ where $\Delta E$ is a regularized gravitational self-energy difference associated to a macroscopic superposition of mass densities.

    Penrose's arguments for gravitational decoherence are not model-specific, but his proposal for the decoherence timescale is the same with the one obtained from Diosi's model, hence the name Diosi-Penrose for this theory.

 Diosi's model predicts a small violation of energy conservation, which can be used to test the theory. A recent experiment rules out the most natural version of the model where $\Lambda$ is the nuclear scale \cite{DPC}. There is no obvious physical justification for the smaller value of the cut-off.

There are also some theoretical problems in the D-P theory. A recent analysis \cite{ALS} of the quantization of weak gravity interacting with matter has shown that the ambiguity pointed out by Penrose is not related to the gravitational self-energy. The latter appears the same in all gauges / reference frames. The frame-dependent terms on the Hamiltonian are different, and they have no obvious physical interpretation.

Moreover, the Diosi master equation   implies a coupling of quantum matter to gravity that differs from that of GR,  in effect, through a non-unitary channel. This channel should also persist at the macroscopic level (for example, via non-conservation of energy). As a result, the resulting theory for matter interacting with gravity cannot be a Hamiltonian theory, hence, it cannot be GR. Even if the avowed aim of Penrose is to gravitize quantum mechanics, this specific approach apparently requires strong modifications in our current theory of gravity, and not only in quantum mechanics.

\subsubsection{The Newton-Schr\"odinger equation}
\label{NSE}

An early candidate for the coupling of classical gravity to quantum matter is the so-called Moller-Rosenfeld (MR) theory \cite{Moller, Rosenfeld},  according to which the source in Einstein equations in the expectation value of the stress-energy tensor:   $G_{\mu \nu} = \frac{\kappa}{2} \langle\hat{T}_{\mu \nu}\rangle$  is taken as fundamental in this theory, rather than an approximation to a quantum gravity theory. (See Ref. \onlinecite{HuSCG} for how  the MR theory, at least the way it is interpreted by recent followers,  differs from semiclassical gravity theory proper~\cite{HuVer20},  based solely on GR+QFT).    It is far from obvious that the MR theory is consistent, or even that it makes mathematical sense, but at least in the weak field, nonrelativistic regimes, one can subject it to tests against theories based on GR+QFT. In these regimes, adherents of the M-R theory like to work with  the Newton-Schr\"odinger equation (NSE) \cite{NS}, which is a non-linear equation for the wave-function
\begin{eqnarray}
i\frac{{\partial}\psi}{{\partial t}}  = - \frac{1}{2m} \nabla^2 \psi + m^2 V_N[\psi]  \psi  \label{NS}
\end{eqnarray}
where 
$V_N(\bf r)$ is the (normalized) gravitational (Newtonian) potential given by
\begin{eqnarray}
V_N({\bf r},t) = - G \int d{\bf r'} \frac{|\psi({\bf r'},t)|^2 }{|{\bf r} - {\bf r'}|}.
\end{eqnarray}
Note, as we have emphasized in Ref. \onlinecite{AHNSE}, such a single (N=1) particle NS equation is not derivable from GR+QM. It is emphatically not a representative of semiclassical gravity based on GR and QFT, which can be viewed as the large N limit of quantum gravity \cite{HarHor}.

\subsection{Decoherence from quantum gravity processes}

This involves proposals about decoherence from Planck scale processes that cannot be identified by low energy physics. Early models attempted to find a connection with non-unitarity suggested by the black hole information paradox \cite{NaSr84}. Now, we are well aware that there is no direct relation: the presumed non-unitarity in black hole evaporation is accompanied with the loss of Cauchy surfaces, hence, with the loss of the global notion of a state. Therefore, it is incompatible also with the non-unitarity of Markovian master equations that describe the evolution of a globally defined state \cite{InfLo}.

Gravitational decoherence of this type is presumed to originate from specific features of quantum gravity theories, for example, string-theory couplings to massive-string states \cite{EMN92} or emergent non-locality from spacetime foam \cite{Garay98}. The key critique of this type of models is presented in Ref. \onlinecite{AHGravDec}.

Any entity of the quantum gravity realm, such as spacetime foam,  exists at the Planck scale,  before spacetime with a Lorentz structure emerges. One needs a strong case to show
that some Planck-scale properties can escape the coarse-graining and scaling which subsumes their effect to that of the average as the large scale manifold structure of spacetime emerges, and emerge at low energies. The most reasonable assumption is that their average behavior is contained in the effective description that survives at the low energy
limit, namely, GR; we can think of no counter-example in all fields of physics. Even if some Planck-scale effects survive at low energies, they must be much smaller than the effects that originate from the gravitational dynamics of GR. Hence, decoherence from quantum gravity effects will be subdominant compared to all decoherence effects that originate from GR, or are postulated on the basis of GR.

\subsection{Decoherence from Spacetime fluctuations}
\label{spctfl}
 In these models decoherence originates from   some fundamental imprecision in the measuring devices (starting with clocks and rulers) \cite{GPP} or uncertainties in the dynamics \cite{MilIntDec}, or treating time as a statistical variable \cite{Bonifacio}. Usually, the fluctuations are assumed to originate from Planck-scale physics.

These approaches make very strong assumptions about the physics of the Planck scale. There is absolutely no reason to expect that such uncertainties can be modeled by stochastic processes which are  intrinsically classical, as is commonly the case in this approach. The modeling of  uncertainties by classical noise is a valid assumption for randomness at the macroscopic scale.  In contrast, quantum uncertainties are different in nature  as they involve non-localities and correlations with no analogues in the classical theory of stochastic processes.

To explain this point, we note that the limitations posed by the Planck length are not {\em a priori} different from those placed by the scale $\sqrt{\hbar/c^3} q/m$ in quantum electrodynamics: At this scale quantum field effects are strong, and the fluctuations from these effects are fully quantum. Any effect they
cause at low energy is also inherently quantum. One needs to  specify the conditions (e.g., Gaussian systems) or the regime for the quantum field,  and justify the means   by which they could be  treated like classical fluctuations described by a stochastic process. In particular, the effects of the fluctuations of the EMF at low energies ($E<<mc^2$) have been well studied. It has been shown that the `noise' induced by these fluctuations is non-Markovian and does not cause significant decoherence effects in the microscopic regime \cite{Emdec}.  In other words, the coherence of the EMF vacuum does not allow for the {\em a priori} generation of classical (i.e., decohering) fluctuations in the quantum motion of the particle.  The assumption that the gravitational field exhibits a different behavior is completely {\em ad hoc}, with no justification unless one postulates that gravity is fundamentally classical.

Note that the theory of semiclassical stochastic gravity (ScStG) \cite{HuVer20} can introduce stochasticity
associated solely with the quantum fluctuations of matter fields. The backreaction of the mean and the fluctuations of the stress energy tensor of a quantum matter field is taken into account by way of the Einstein-Langevin equation \cite{ELE} governing the dynamics of  the {\it induced} metric fluctuations. Simple dimensional analysis shows that these induced metric fluctuations are dominant only at the Planck scale, and thus are on the same footing as  spacetime foams, albeit Wheeler wanted spacetime foams to admit topology changes. In ScStG the noise is fundamentally quantum, they are not put in by hand, and the theory is entirely based on GR +QFT. The drop off behavior of metric fluctuations in Minkowski spacetime has been derived in  \onlinecite{MarVer}.  Indeed,  the strength of quantum noise in ScStG provides a useful measure of the validity of semiclassical gravity\cite{HRValidSCG}).  In contrast, the noise employed by the  models discussed above is defined by a purely classical stochastic process.

A second problem in these approaches is that they often contradict the symmetries of GR.  Fluctuations in the time or space coordinates of an event are indistinguishable mathematically from time and space reparameterizations of the system, only such reparameterizations are viewed as stochastic. However, time and space reparameterizations are pure gauge variables in classical relativity (even in the linearized approximations); they do not have any dynamical content. The invariance of the theory under space and time reparameterizations follows from the diffeomorphism invariance of the classical action, a fundamental symmetry of general relativity.

The assignment of dynamical contents to such fluctuations implies that they are not treated as gauge variables.
Doing so violates the fundamental symmetry of classical GR.   Any theory with this property would have far-reaching implications which goes beyond the gravitationally-induced decoherence effects. The diffeomorphism symmetry affects both the dynamics and the kinematics of general relativity, and its abandonment ought to be manifested in other gravitational phenomena.

We also note that time and space reparameterizations
decouple from the terms describing Newtonian interaction, already at the classical level. Hence, there is no reason for  Newton’s constant in GR to modulate the strength of decoherence effects.

We may also place in this category the so-called {\em event operator formalism} of Ralph, Milburn and Downes \cite{event1, event2}. This model originates from a modification of standard QFT that allows for unitary evolution in presence of closed timelike  curves. This model does not lead to decoherence of individual particles, but it  predicts rather strong decorrelation of entangled photon pairs.

The presumed physics in this model is rather implausible from the perspective of gravity theory. The event operator formalism relies on the presence of closed-timelike curves \cite{deutsch} without accompanying quantum gravity effects. This strongly contradicts the well-motivated  {\em chronology protection conjecture} \cite{HawkingCP} which asserts that the laws of physics, including quantum phenomena, do not allow for the appearance of closed time-like curves. Chronology protection is valid also for semi-classical gravity \cite{KRW}. All this  strongly suggests  that closed timelike curves can emerge only as Planck scale quantum gravity effects (like Wheeler's spacetime foam \cite{Wheeler}), if at all.


\section{ABH theory:  based on GR and QM}

\subsection{Key ideas and results}
In the theory of gravitational decoherence of  Anastopoulos \& Hu \cite{AHGravDec} and Blencowe \cite{Miles}  (ABH) the source of decoherence comes as noise  (fluctuations) from gravitational waves (classical perturbations) or gravitons (quantized linear perturbations), or simply, metric fluctuations.
These fluctuations satisfy Einstein's equations and as such they are to be identified with transverse-traceless perturbations. The source of these fluctuations may be cosmological \cite{Miles} (stochastic gravitons produced in the early universe near the big bang or from inflation pre- or post), astrophysical \cite{ReJa} or structural to GR as an emergent theory---see Ref. \onlinecite{E/QG} for an explanation. Note also the later work, Ref. \onlinecite{Oniga}, which  derives a related theory of gravitational decoherence for matter and light.

  There is one parameter in the master equation of the ABH theory which is the noise temperature $\Theta$.  It coincides with the graviton temperature if the origin of perturbations is cosmological. Early works in this direction includes the study of decoherence due to the graviton vacuum \cite{Ana96}, and of open system dynamics of particles in  graviton baths \cite{HaKl, Rey}. The Power-Persival \cite{PowPer} collapse model  also falls in the same category, but the  perturbations there are restricted to conformal waves.

It is important to distinguish the ABH theory from models of gravitational decoherence from metric perturbations that do not require the perturbations to satisfy the linearized Einstein equations. Such models have been  developed by Kok and Yurtsever \cite{KoYu};  Breuer, G\"okl\"u and L\"ammerzahl  \cite{Breuer};   and   Asprea,  Gasbari, Ulbricht and Bassi \cite{AGB}. These models describe the perturbations of the metric  by a stochastic process, but they do not distinguish between true and pure gauge degrees of freedom in the perturbations. Treating variables that are pure gauge in GR  as stochastic  is not compatible with GR, because they do not implement the diffeomorphism symmetry. (Unless one assumes   a stochastic matter source of unknown origin and physics, to which the pure gauge stochastic perturbations are slaved.) They are intermediate between the ABH model and the models of Sec. \ref{spctfl}, in that they postulate stochastic behavior of pure-gauge quantities, but they also include a contribution from true degrees of freedom.

The ABH  type of analysis  applies to any type of quantum matter fields in addition to scalar, any number of particles, in the ultra-relativistic as well as the non-relativistic regimes.  The specific methodology is the following.

The system under consideration is a quantum field (a massive scalar field in the simplest case), interacting with a gravitational field as its environment.
Gravity is described by classical general relativity.  In the weak field limit, we describe gravitational perturbations in the linearized approximation. Hence, one starts
   from a linearization of  the Einstein-Hilbert action around Minkowski spacetime,
and constructs the associated Hamiltonian through a 3+1 decomposition. The constraints of the system are solved classically, thereby allowing us to express the Hamiltonian
in terms of the true physical degrees of freedom of the theory, namely,
the transverse-traceless perturbations for gravity and the scalar field.
We then quantize the scalar field and the gravitational perturbations and trace-
out the contribution of the latter.

A key input in this stage is the specification of
an initial state for the gravitational perturbations. We consider an initial condition
that interpolates between the regime of negligible (vacuum) perturbations and strong
classicalized perturbations. The initial state is defined in terms of a free parameter $\Theta$
that can be loosely interpreted as the noise temperature of the perturbations. $\Theta$    conveys coarse-grained information reflective of the micro-structures of spacetime, similar to temperature with regard to molecular motion, or the  spectral density function of the environment in Brownian motion.  It is in this sense that we think gravitational decoherence may reveal the underlying ‘textures’ of spacetime beneath that described by classical general relativity.

Following the standard methodology of open quantum systems a  2nd order (perturbative) master equation for the matter
field is obtained.  This master equation applies to configurations with any number of particles.
We project the master equation to the single-particle subspace and we derive a master equation for a single particle. The latter simplifies significantly in the non-relativistic
regime,  leading to the ABH master equation
\begin{eqnarray}
 \frac{\partial \hat{\rho}}{\partial t} = -i [\hat{H}, \hat{\rho}] - \frac{\tau}{16m^2} (\delta^{ij} \delta^{kl} + \delta^{ik}\delta^{jl}) [\hat{p}_i
\hat{p}_j,[\hat{p}_k\hat{p}_l, \hat{\rho}]] \label{ABHME}
\end{eqnarray}
 where $\tau$ is a constant of dimension time and $\hat{H} = \frac{\hat{p}^2}{2m}$.

 For motion in one dimension, the ABH master equation simplifies to
 \begin{eqnarray}
 \frac{\partial \hat{\rho}}{\partial t} = -i [\hat{H}, \hat{\rho}] - \frac{\tau}{2} [\hat{H},[\hat{H},\hat{\rho}]], \label{1dim}
 \end{eqnarray}
 where $\tau$ is a constant of dimension time and $\hat{H} = \frac{\hat{p}^2}{2m}$.

 In the ABH model
 \begin{eqnarray}
 \tau = \frac{32\pi G \Theta}{9} = \frac{32\pi}{9} \tau_P (\Theta/T_P),
 \end{eqnarray}
 where $T_P = 1.4\times 10^{32} \; K$ is the Planck temperature.

 The   master equation (\ref{1dim}) appears in models by Milburn \cite{MilIntDec}, Adler \cite{Adler}, Diosi \cite{Diosi}and Breuer et al \cite{Breuer} where $\tau$ is obtained from postulated stochastic fluctuations of time, discreteness of time, stochastic fluctuations of the metric,  or even stochastic fluctuations of $\hbar$. In these models, the value of $\tau$ is not fixed, but the natural candidate is the Planck-time $\tau_P = 10^{-43}s$.  In Ref. \onlinecite{AHGravDec}, it is argued that $\tau$ need not be restricted to  $ \tau_P$  but  can be  a free parameter depending on the underlying structures of spacetime at different scales which may vary in different theories.

The ABH model can be generalized to photons, where we can obtain a master equation for a general photon state \cite{LagAn}. For a single photon,
\begin{eqnarray}\label{result}
 \frac{\partial \hat{\rho}}{\partial t} &= -i \, [ \hat{H } , \hat{\rho}] - \frac{\tau_{ph}}{2}   \left( \delta^{in} \delta^{jm} -\frac{1}{3} \delta^{ij} \delta^{nm} \right) \left[ \frac{\hat{p}_i \, \hat{p}_j}{\hat{p}_0} , \left[ \frac{\hat{p}_n \, \hat{p}_m}{\hat{p}_0}, \hat{\rho}  \right] \right] \, ,
\end{eqnarray}
where $\hat{H} = |\hat{\bf p}|$ and $\tau_{ph} = 4 G \Theta$.

\subsection{Observational constraints to the ABH}

If we regard the parameter  $\Theta$ in the ABH theory  as a noise temperature---originating from some emergent gravity theory---, then $\Theta$  need not be related to  the Planck length and even $\Theta >> T_P$ is  perfectly acceptable from a theoretical point of view. $\Theta$ is only a measure of the power $P$ carried by the noise,
 $P \sim\Theta \Delta \omega$, where $\Delta \omega$ is the band-width of the noise.

Some bounds to $\Theta$ can be estimated from the non-relativistic analysis of \onlinecite{Ana96}. This paper employs the Feymnan-Vernon influence functional method, which has the benefit of providing a simple stochastic equation for the semi-classical evolution of a particle interacting with a heat bath. The results that we will present are new, but they follow from the direct use of the `thermal' noise  of the ABH model to the analysis of Ref. \onlinecite{Ana96}.

The effective semiclassical equation for a non-relativistic particle in presence of classical gravitational perturbations of noise temperature $\Theta$ turns out to be
\bey
\ddot{x} + \frac{2G}{15} \ddot{x}^2 \dot{x} = 2 \ddot{x}\xi \label{noise}
\eey
where $\xi(t)$ is Gaussian noise with $\langle \xi(t) \xi(t')\rangle = \eta(t-t')$, where $\eta$ is known as the noise kernel. The dissipative term is relatively weak at it corresponds to energy loss due to gravitational wave radiation, so we can ignore it.

The noise kernel for the ABH model is
\bey
\eta(s) = \frac{G}{2} \int_0^{\Lambda} dk k \cos(ks) \coth\frac{k}{2\Theta},
\eey
where $\Lambda$ is a cut-off and $\Theta$ the noise temperature. The physically relevant regime corresponds to $\Theta >> \Lambda$, whence,
\bey
\eta(s) = \pi G\Theta \delta (s). \label{noisek}
\eey
By Eq. (\ref{noise}),  the noise behaves like  a  stochastic fluctuations of the particle's inertial  mass of order
\bey
\frac{\delta m}{m} \sim \sqrt{\xi^2(t)} \label{heuris}
\eey
With the noise kernel (\ref{noisek}) we find
\bey
\frac{\delta m}{m} \sim \sqrt{\frac{\Theta}{T_P}}, \label{deltamG}
\eey
where $T_P$ is the Planck length.
We can use Eq. (\ref{deltamG}) to establish bounds to $\Theta$ from cosmological and solar-system measurements.

Cosmology does not lead to strong constraints in $\Theta$. If we assume that the $\Lambda$CDM model holds we can identify the maximum value of $\frac{\delta m}{m}$ with  the relative error in the determination of baryon mass density, hence $\frac{\delta m}{m} \sim 10^{-2}$, which implies that $\Theta < 10^{-4} T_P$. However, if we take into account the changes in the values of the baryon mass density in different dark energy models, a bound $\frac{\delta m}{m} \sim 10^{-1}$ is more plausible, but even this may be too restrictive. We need a model that includes intrinsic stochastic gravitational perturbations in the evolution of the Universe, in order to estimate a proper bound to the size of these fluctuations.

Solar system measurements provide a better constraint to $\Theta$, if we assume that Eq. (\ref{deltamG}) also applies to large astronomical bodies like planets. This assumption is by no means evident, because the derivation of Eq. (\ref{noise}) treats particles as pointlike, or at least, much smaller than the typical wavelength of gravitational perturbations. Eq. (\ref{deltamG}) overestimates the effects of the gravitational noise. In an extended system, part of the noise would be expended on the moment of inertia and on higher moments  of the mass density, leading to weaker effects on the center of mass motion.

In any case, if we take Eq. (\ref{deltamG}) to apply to planets, we can use the relative accuracy in the measurement of Earth's mass to place an upper bound $\frac{\delta m}{m} < 10^{-4}$, which implies that $\Theta < 10^{-8} T_P$.  This estimate is probably too stringent, because the  most accurate measurements of  mass in astronomical bodies comes from the measurement of the gravitational acceleration on its surface, not from its orbit in the solar system, as would be required for comparison with Eq. (\ref{noise}). An exact bound on $\Theta$ will require an analysis of the motion of planet-sized bodies under ABH-type noise. However, our simple analysis shows that under pessimal assumptions, sufficiently large values of $\Theta$ that are compatible with observable decoherence effects cannot be ruled out on the basis of existing measurements.

\section{Types of experiments}

In this section, we provide order-of-magnitude estimates for ABH-type decoherence, in comparison with the D-P model.

\subsection{Optomechanical experiments}

Consider a body brought into a superposition of a zero momentum and a finite momentum state, corresponding to an energy difference $\Delta E$. For the ABH model, the decoherence rate for the center of mass is then
     \begin{eqnarray}
     \Gamma_{ABH} = \frac{ (\Delta E)^2 \tau}{\hbar^2},
     \end{eqnarray}
     where $\tau$ is the free parameter in the master equation (\ref{ABHME}). A value for $\Gamma_{ABH}$ of the order of $10^{-3}s$ may be observable in optomechanical systems, as it is competitive with current environment-induced-decoherence timescales. Hence,  to exclude values of $\tau > \tau_P$, we must prepare a quantum state with  $\Delta E  \sim 10^{-14}$J.

In the DP model, the decoherence rate for a sphere of mass $M$ of radius $R$ in a quantum superposition of states with different center of mass position (though the predicted decoherence rate is largely independent of the details of the prepared state) is of the order of
     \begin{eqnarray}
     \Gamma_{DP} = \frac{GM^2}{\hbar \sqrt{R^2+\ell^2}},
     \end{eqnarray}
     where $\ell$ is a cut-off length, originally postulated to be of the order of the size of the nucleus, but recently constrained to $\ell > 0.5 \cdot 10^{-10}m$ (see Ref. \onlinecite{DPC}).  Alternative models postulate $\ell$ up to a scale of $10^{-7}$m. For an optomechanical nanosphere with $M \sim 10^{10}$ amu and $R \sim 100$ nm, $\Gamma_{DP} \sim 10^{-3}s^{-1}$, a value that is in principle measurable in optomechanical experiments.

\subsection{ Matter wave interferometry.} In far field interferometry, the ABH model (but not the 1-d master equation (\ref{1dim}))  leads to loss of phase coherence of the order of
 $(\Delta \Phi)^2 = m^2 v^3 \tau L/\hbar^2$, where $L$ is the propagation distance inside the interferometer. While the exact derivation of $(\Delta \Phi)^2$ requires a dynamical analysis, it  is of the order of $\Gamma_{ABH} t_{int}$, where $t_{int} = L/v$ is the average time of the particle in the interferometer.
 Setting an upper limit of $L = 100$ km, and $v = 10^4$ m/s, decoherence due to cosmological gravitons requires particles with masses of the order of $10^{16} $amu. If $\Theta$ is a free parameter, experiments with particles at $10^{10}$amu
 will test up to $\Theta \sim 10^{-5}T_P$. For comparison,  the heaviest molecules used to date in quantum mechanical interference experiments are oligoporphyrines with mass of ``only'' $2.6 \cdot 10^4\,$amu \cite{Fein19}.
  ftfc🙌
 The Diosi-Penrose model and other models that lead to decoherence in the position basis can also be tested by near-field \cite{MAQRO} and far-field \cite{MAQRO0} matter-wave interferometry.
A rough estimation for the loss of phase coherence is $ (\Delta \Phi)^2  \simeq \Gamma_{DP}L/v = \frac{Gm^2L}{\hbar Rv}$, where $R$ is the radius of the particles.  In contrast to the ABH model, this loss of coherence is enhanced at low velocities. Assuming $L = 100$ km, $v = 10$ m/s, and $R = 100$ nm, an experiment would require a mass $M \sim 10^9-10^{10}$ amu to observe decoherence according to the DP model.

\subsection{Wave-packet spread.} The intrinsic spreading of a matter wave-packet in free space is a hallmark of Schr\"odinger evolution. ABH-type models predict negligible deviations in the wave-packet spread from that of unitary evolution. The DP model and all other models that involve decoherence in the position basis predict
 a wave packet spread of the form
  \begin{eqnarray}
(\Delta x)^2(t) = (\Delta x)_S^2(t) + \frac{\Lambda}{2m^2} t^3,
\end{eqnarray}
where $(\Delta x)_S^2(t)$ is the usual Schr\"odinger spreading, and $\Lambda$ depends on the model. The changes from free Schr\"odinger evolution become significant at later times.  An exact estimation of this effect depends on properties of the initially prepared state, and  is rather involved. The MAQRO proposal \cite{} estimates that for a free-propagation time equal to $100$s  (accessible in their setup) it is possible to constrain CSL-type models, some models of quantum gravity decoherence, but not decoherence of the D-P type.

In contrast, the Newton-Schr\"odinger Equation predicts a {\it retraction} of the wave-packet spread   for masses around $10^{10}$ amu \cite{GiGr}. An osmium nanosphere of radius $R \simeq 100$ nm would require a couple of hours of free propagation in order to observe significant deviation from Schr\"odinger spreading \cite{Grossa}.

\subsection{Decoherence of photons} Only the ABH model has been generalized for photons \cite{Oniga, LagAn}. For interferometer experiments with arm length $L$, the model predicts loss of visibility of order $(\Delta \Phi)^2 = \frac{8G\Theta E^2L}{\hbar^2 c^6}$. For $L=10^5$km, $\Theta \sim T_P$ and photon energies $E$ of the order of 1eV, this implies a loss of coherence of the order of $\Delta \Phi = 10^{-8}$.  In principle, this would be discernible with EM-field coherent states with mean photon number $\bar{N} > 10^{16}$, though it would be very challenging to suppress all other systematic errors to this degree.

The linear  dependence of $\Delta \Phi$ on energy implies that decoherence is significantly stronger at high
frequencies. For interferometry in the extreme UV, $\Delta \Phi$  may increase by two orders of
magnitude or more.
     Alternative set-ups, such as the formation of effective Fabry-Perot `cavities' with mirrors could increase the effective propagation length by many orders of magnitude, and hence, lead to stronger signatures of ABH-predicted, photon gravitational decoherence.

\section{Conclusion}

  Quantum gravity research, the quest for theories of the microscopic constituents of spacetime, in achieving the fusion of quantum and gravity (Q$\times$G) at the Planck scale ($10^{-35}m$),  has occupied the attention of a significant number of theoretical physicists for the past seven decades. Yet the lack of  observable experimental data has prevented any of the resulting theories to claim success.  Instead of chasing this lofty yet unattainable goal, with little chance of finding directly verifiable evidence at today's energy,  we should set our targets at the union of quantum and gravity (Q+G) because the contradictions and inconsistencies between  quantum theory and general relativity (GR) already show up acutely at today's accessible low energy scales and there are earthbound and space experiments which can probe into issues at the joining  of these two fundamental theories.

  In particular, taking advantage of long baselines and small environmental influence, deep space experiments \cite{DSQL,expdec, MAQRO, MAQRO0}  can provide significant novel information about the coexistence of quantum and gravity -- no matter how precarious it is -- and separate them from the alternative theories .  In this context gravitational decoherence experiments  have  far-reaching theoretical significance in at least two respects: i) discriminating alternative  quantum or gravity theories, such as those mentioned earlier, based on their predictions,  against that from theories based on general relativity and quantum field theory, such as the ABH theory described above; ii)  the possibility of discerning the nature of gravity, whether it is a fundamental theory or an effective theory.

Concerning the first aspect,  it is easy to make the demarcation because most proposals for gravitational decoherence involve a violation  of quantum theory, or of GR, or (usually) of both, whereas the ABH theory respects both QFT and GR.  Precision experimental data can quickly discern these two categories of theories.  Concerning the second aspect,  why it is important and how can one make such a distinction, we offer some background perspective in the following.

GR is commonly accepted as the best theory for the description of macroscopic spacetime, but whether quantizing GR will yield the true theory of the microscopic structure of spacetime  at the Planck scale remains an open question.  GR could well be  an effective theory emergent from some fundamental theory of  quantum gravity,   valid only at the macroscopic scale we are familiar with.  The ABH model  distinguishes these two alternatives.  In the fundamental  theory view,  Minkowski spacetime is  the ground state of a quantum gravity theory.  In the emergent theory view, Minkowski spacetime is a low energy collective state or  {\em macrostate} of quantum gravity, whereby one could associate a thermodynamic  description. The key difference between a ground state and a macrostate is not energy,  but the {\em strength of fluctuations}: thermodynamic fluctuations are much stronger than quantum  fluctuations in spacetime, and they can cause significantly stronger decoherence.

Therefore, if we can see evidences of  gravitational decoherence, then, using a theory based on GR+QFT such as the ABH theory,   even crude orders of magnitude differences in the observation data could provide a strong discriminant separating gravity as a fundamental theory from an effective one.    \\

\noindent {\bf Acknowledgment}  On the theoretical side, the ABH theory presented here refers to work done not only by us but also independently by Miles Blencowe, with whom we enjoy synergetic interactions.  The experimental considerations presented here are assimilated in the grand review paper produced by the  DSQL team led and coordinated by Makan Mohageg of JPL, whose persistent interest in the ``texture of spacetime" served as an stimulant to our investigations.  This work is supported in part by a Schwinger foundation grant JSF-19-07-0001;  BLH is supported in part by NASA/JPL grant 301699-00001.


\newpage
\begin{thebibliography}{00}

\bibitem{Karol} F. Karolyhazy, {\em Gravitation and Quantum Mechanics of Macroscopic Objects }, Nuovo Cim. 52, 390 (1966).

\bibitem{Diosi0} L. Diosi, {\em A Universal Master Equation for the Gravitational Violation of Quantum Mechanics}, Phys. Lett. A 120, 377 (1987).

  L. Diosi, {\em   Models for Universal Reduction of Macroscopic Quantum Fluctuations}, Phys. Rev. A40, 1165 (1989).

\bibitem{Penrose} R. Penrose, {\em Gravity and State Vector Reduction},  in Quantum Concepts in Space and Time, R. Penrose and C. J. Isham editors, (Oxford,
1986, Clarendon Press)

R. Penrose, {\em On Gravity's role in Quantum State Reduction},  Gen. Rel. Grav. 28, 581 (1996).

\bibitem{AnHu15} C. Anastopoulos and B. L. Hu, {\em Probing a Gravitational Cat State}, Class. Quant. Grav. 32, 165022 (2015).


\bibitem{Griffiths} R. B. Griffiths, {\em Consistent Histories and the Interpretation of Quantum Mechanics}, J. Stat.  Phys., 36, 219 (1984).

R. B. Griffiths, {\em Consistent Quantum Theory} (Cambridge University Press, Cambridge 2002).

\bibitem{Omnes} R. Omn\'es, {\em  Logical Reformulation of Quantum Mechanics. I. Foundations},  J.  Stat. Phys. 53, 893 (1988).

R. Omn\'es,  {\em Consistent Interpretations of Quantum Mechanics}, Rev. Mod. Phys.  64, 339 (1992).

 R. Omn\'es, {\em The Interpretation of Quantum   Mechanics}, (Princeton University Press, Princeton 1994).

\bibitem{GeHa}  M. Gell-Mann and J. B. Hartle, {\em Quantum Mechanics in the Light of Quantum Cosmology}, in ‘Complexity, Entropy, and the Physics of Information’, ed. W. Zurek, (Addison Wesley, Reading 1990).

  M. Gell-Mann and J. B. Hartle,  {\em Classical Equations
for Quantum Systems},  Phys. Rev.   D47, 3345 (1993).

 J.B. Hartle,
{\em Spacetime Quantum Mechanics and the
Quantum Mechanics of Spacetime}
in `Gravitation and Quantizations',  in the
Proceedings of the 1992 Les Houches Summer School, ed. by B. Julia and J. Zinn-
Justin,  Les  Houches  Summer  School  Proceedings,
Vol. LVII, (North Holland, Amsterdam, 1995); [gr-qc/9304006].


\bibitem{KeDo} F. Dowker and A. Kent, {\em On the Consistent Histories Approach to Quantum
Mechanics}, J. Stat. Phys., 82, 1575 (1996).

\bibitem{DHmeas}E. Okon and D. Sudarsky, {\em Measurements according to consistent histories}, Stud. His. Phil. Mod. Phys.  48, 7  (2014).

 R. B. Griffiths, {\em Consistent Quantum Measurements},  arXiv:1501.04813.

 E. Okon and D. Sudarsky,{\em The Consistent Histories Formalism and the Measurement Problem}, Stud. Hist. Phil. Mod. Phys. 52, 217 (2015).

\bibitem{Zurek} W. H. Zurek, {\em  Environment-Induced Superselection Rules} Phys. Rev. D26,
1862 (1982).

\bibitem{JoZe} E. Joos and H. D. Zeh, {\em
The Emergence of Classical Properties through
Interaction with the Environment}, Zeit. Phys. B59, 223 (1985).

\bibitem{Zeh} H. D. Zeh, {\em  On the Interpretation of Measurement in Quantum Theory}, Found.
Phys. 1, 69 (1970).

\bibitem{Schwinger} J. S. Schwinger, {\em Brownian Motion of a Quantum Oscillator}, J. Math. Phys. 2, 407 (1961).

\bibitem{FeVe} R. P. Feynman and F. L. Vernon, {\em The Theory of a General Quantum System Interacting with a Linear Dissipative System},
Ann. Phys. 24, 118 (1963).

\bibitem{CaLe} A. O. Caldeira and A. J. Leggett, {\em  Path Integral Approach to Quantum Brownian Motion}, Physica 121A,  587 (1983).



\bibitem{HPZ}   B. L. Hu, J. P. Paz, and Y. Zhang, {\em Quantum Brownian Motion in a General Environment: Exact Master Equation with Nonlocal Dissipation and Colored Noise}, Phys. Rev. D45, 2843 (1992).


\bibitem{decohmeas} C. Anastopoulos, {\em Frequently Asked Questions about Decoherence}, Int. J. Theor. Phys. 41,  1573 (2002).

M. Schlosshauer, {\em Decoherence, the Measurement Problem, and Interpretations of Quantum Mechanics}, Rev. Mod. Phys. 76,  1267 (2004).

M. Schlosshauer, {\em Quantum Decoherence},  	Phys. Rep. 831, 1 (2019).


\bibitem{GRW} G. C. Ghirardi, A. Rimini and T. Weber, {\em Unified dynamics for microscopic
and macroscopic systems}, Phys. Rev.
D34, 470 (1986).

\bibitem{Pearle}
P. Pearle,  {\em Reduction of the State Vector by a Nonlinear Schrödinger Equation}, Phys. Rev. D, 13, 857 (1976).

P. Pearle, {\em Toward Explaining Why Events Occur}, Int. J.  Theor. Phys. 18, 489 (1979).

P. Pearle, {\em Combining Stochastic Dynamical State-Vector Reduction with Spontaneous Localization}, Phys. Rev. A39, 2277 (1989).

 \bibitem{various}  G. C. Ghirardi, P. Pearle and A. Rimini, {\em Markov Processes in Hilbert Space and Continuous Spontaneous Localization of Systems of Identical Particles}, Phys. Rev. A42, 78 (1990).

G. C. Ghirardi and A. Bassi, {\em  Dynamical Reduction Models}, Phys. Rep. 379, 257 (2003).

A. Bassi, K. Lochan, S. Satin, T. P. Singh, and H. Ulbricht, {\em Models of Wave-Function Collapse, Underlying Theories, and Experimental Tests}, Rev. Mod. Phys. 85, 471 (2013).


\bibitem{CSLQFT} R. Tumulka, {\em A Relativistic Version of the Ghirardi–Rimini–Weber Model}, J. Stat. Phys.  125, 821 (2006).

    R. Tumulka, {\em  On Spontaneous Wave Function Collapse and Quantum Field Theory}, Proc. Roy. Soc. A462, 1897 (2006).

D. J.  Bedingham,  {\em Relativistic State Reduction Dynamics}, Found.  Phys., 41, 686 (2011).

D. Bedingham, D. D\"urr, G. C. Ghirardi, S. Goldstein, R. Tumulka, and N. Zanghì,  {\em Matter Density and Relativistic Models of Wave Function Collapse}, J. Stat. Phys., 154, 623 (2014).

P. Pearle, {\em Relativistic Dynamical Collapse Model}, Phys. Rev. D91, 105012 (2015).

\bibitem{JGB} C. Jones, G. Gasbarri and  Angelo Bassi, {\em
Mass-coupled relativistic spontaneous collapse models}, J. Phys. A: Math. Theor. 54, 295306 (2021).

\bibitem{gravdec} A. Bassi, A. Großardt, and H. Ulbricht, {\em Gravitational Decoherence},  	Class. Quant.  Grav. 34, 193002 (2017).



\bibitem{Penrose2} R. Penrose,  {\em On the Gravitization of Quantum Mechanics 1: Quantum State Reduction},  Found. Phys.44, 557 (2014).


\bibitem{AHGravDec} C. Anastopoulos and B. L. Hu,
 {\em A Master Equation for Gravitational Decoherence: Probing the Textures of Spacetime},  Class. Quant. Grav. 30, 165007 (2013)


\bibitem{Miles} M. Blencowe, {\em Effective Field Theory Approach to Gravitationally Induced Decoherence}, Phys. Rev. Lett. 111, 021302 (2013).  	


\bibitem{dephasinggravity}  I. Pikovski, M. Zych, F. Costa, and C. Brukner, {\em Universal Decoherence due to Gravitational Time Dilation}, Nat. Phys. 11, 668 (2015).

  I. Pikovski, M. Zych, F. Costa, and C. Brukner, {\em Time Dilation in Quantum Systems and Decoherence}, New J. Phys. 19,  025011 (2017).

\bibitem{hahn} E. L. Hahn, {\em Spin Echoes}, Phys. Rev. 80, 580 (1950).

\bibitem{AnSav11} C. Anastopoulos and N. Savvidou,  {\em Consistent Thermodynamics for Spin Echoes}, Phys. Rev. E83, 021118 (2011).





\bibitem{DPC}S. Donadi, K. Piscicchia, C. Curceanu, et al, {\em Underground Test of Gravity-Related Wave Function
Collapse}  Nat. Phys. (2020).https://doi.org/10.1038/s41567-020-1008-4


\bibitem{ALS} C. Anastopoulos, M. Lagouvardos and K. Savvidou, {\em Gravitational effects in macroscopic quantum systems: a first-principles analysis},  Class. Quantum Grav. 38, 155012 (2021).

\bibitem{Moller} C. Moller, {\em Les Theories Relativistes de la Gravitation}, "Colloques Internationaux CNRX 91", edited by A Lichnerowicz and M.-A. Tonnelat (CNRS, Paris 1962).

\bibitem{Rosenfeld} L. Rosenfeld, {\em On Quantization of Fields }, Nucl. Phys. 40, 353 (1963).


\bibitem{NS} M. Bahrami, A. Großardt, S. Donadi, and A. Bassi, {\em The Schr\"odinger–Newton Equation and its Foundations},  New J. Phys. 16, 115007 (2014).

\bibitem{HuSCG} B. L. Hu, Gravitational Decoherence, {\em Alternative Quantum Theories  and Semiclassical Gravity},   J. Phys. Conf. Ser. 504,  012021 (2014).

\bibitem{HarHor}
J. B. Hartle and G. T.  Horowitz, Ground-state expectation value of the metric in the 1 N or semiclassical approximation to quantum gravity,  Phys. Rev. D24, 257 (1981).

\bibitem{AHNSE}
 C. Anastopoulos and B. L. Hu,   {\em Problems with the Newton-Schrödinger Equations},  New J. Physics  16, 085007  (2014).




\bibitem{NaSr84}D. V. Nanopoulos and M. Srednicki, {\em Search for violations of quantum mechanics},  Nucl. Phys. B 241, 3 (1984).

\bibitem{InfLo} W. G. Unruh and R. M. Wald, {\em Information Loss}, 	
	Rep.  Prog.  Phys. 80,   092002 (2017).

\bibitem{EMN92} J. Ellis, N.E. Mavromatos and  D.V. Nanopoulos, {\em String Theory Modifies Quantum Mechanics},  	Phys. Lett. B293, 37 (1992).

\bibitem{Garay98}  L. J. Garay, {\em Spacetime Foam as a Quantum Thermal Bath}, Phys. Rev. Lett. 80, 2508 (1998).







\bibitem{GPP} R. Gambini, R. Porto and J. Pullin, {\em Realistic Clocks, Universal Decoherence, and the Black Hole Information Paradox}, Phys. Rev. Lett. 93,
240401 (2004).

 R. Gambini, R. Porto and J. Pullin, {\em A relational solution to the problem of time in quantum mechanics and quantum gravity: a fundamental mechanism for quantum decoherence}, New J. Phys. 6,
45 (2004).

\bibitem{MilIntDec} G. J. Milburn, {\em Lorentz invariant intrinsic decoherence}, New J. Phys. 8, 96 (2006).

G. J. Milburn,  {\em Intrinsic decoherence in quantum mechanics}, Phys. Rev. A44, 5401
(1991).

\bibitem{Bonifacio} R. Bonifacio, {\em Time as a Statistical Variable and Intrinsic Decoherence}, Nuovo Cim. B114, 473
(1999)

\bibitem{Emdec} P. M. V. B. Barone and A. O. Caldeira, {\em
Quantum mechanics of radiation damping}, Phys. Rev.
A43, 57 (1991).



C. Anastopoulos and A. Zoupas, {\em Non-equilibrium Quantum Electrodynamics}, Phys. Rev. D58, 105006
(1998).

P. R. Johnson and B. L. Hu, {\em Stochastic theory of relativistic particles moving in a quantum field: Scalar Abraham-Lorentz-Dirac-Langevin equation, radiation reaction, and vacuum fluctuations}, Phys. Rev. D65, 065015 (2002).


\bibitem{HuVer20}
B. L. Hu and E. Verdaguer, {\em Semiclassical and Stochastic Gravity-- Quantum Field Effects on Curved Spacetimes}  (Cambridge University Press, Cambridge, 2020)



\bibitem{ELE}
E. A. Calzetta and B. L. Hu, {\em
Noise and fluctuations in semiclassical gravity}, Phys. Rev. D.49, 6636 (1994).

B. L. Hu and A. Matacz, {\em
Back reaction in semiclassical cosmology: the Einstein--Langevin equation}, Phys. Rev. D51, 1577 (1995).

E. A. Calzetta, A. Campos and E. Verdaguer, {\em
Stochastic semiclassical cosmological models}, Phys. Rev. D56, 2163 (1997).

F. C. Lombardo and F. D. Mazzitelli, {\em
Einstein--Langevin equations from running coupling constants}, Phys. Rev. D55, 3889 (1997).

 \bibitem{MarVer}
R. Mart{\'{\i}}n and E. Verdaguer, {\em
On the semiclassical Einstein--Langevin equation}, Phys. Lett. B 465, 113 (1999).

R. Mart{\'{\i}}n and E. Verdaguer, {\em
Stochastic semiclassical fluctuations in Minkowski spacetime}, Phys. Rev. D61, 1 (2000).
Physical Review D. 2000;61:1--26.



\bibitem{HRValidSCG}
B. L. Hu, A. Roura and E. Verdaguer, {\em
Induced quantum metric fluctuations and the validity of semiclassical gravity}, Phys. Rev. D70, 1 (2004).




\bibitem{event1} T.C.Ralph, G.J.Milburn and T.Downes, {\em Quantum connectivity of space-time and gravitationally induced decorrelation of entanglement},  Phys. Rev. A79,022121 (2009).

\bibitem{event2}T. J. Ralph and J. Pienar, {\em Entanglement decoherence in a gravitational well according to the event formalism},  New J. Phys. 16, 085008 (2014).

 \bibitem{deutsch} D. Deutsch, {\em Quantum mechanics near closed timelike lines}, Phys.Rev.D.44, 3197 (1991)

 \bibitem{HawkingCP} S. W. Hawking, {\em Chronology protection conjecture}, Phys. Rev. D 46, 603 (1992).

\bibitem{KRW} B. S. Kay,  M. J. Radzikowski and R. M.  Wald,  {\em Quantum Field Theory on Spacetimes with a Compactly Generated Cauchy Horizon }, Comm. Math. Phys. 183, 533 (1997).

\bibitem{Wheeler} J. A. Wheeler,   {\em  Geons}, Phys. Rev.97, 511 (1955).

\bibitem{ReJa} S. Reynaud, B. Lamine, A. Lambrecht, P. Maia Neto, and M. T Jaekel, {\em HYPER and Gravitational Decoherence}, Gen.  Relativ.  Gravit. 362271, (2004).


\bibitem{E/QG}  B. L. Hu, {\em Emergent /Quantum Gravity: Macro/Micro Structures Spacetime},  J. Phys. Conf. Ser. 174, 012015 (2009).


\bibitem{Oniga}T. Oniga and C. H.-T. Wang, {\em Quantum Gravitational Decoherence of Light and Matter}, Phys. Rev. D, 93, 044027 (2016).






\bibitem{Ana96} C. Anastopoulos, {\em Quantum Theory of Non-Relativistic Particles Interacting with Gravity}, Phys. Rev.  D54, 1600   (1996).


\bibitem{HaKl} Z. Haba, {\em Decoherence by relic gravitons}, Mod. Phys. Lett. A15, 1519 (2000).


 Z.Haba and H.Kleinert, {\em Quantum-Liouville and Langevin equations for gravitational radiation damping}, Int. J. Mod. Phys. A17, 3729 (2002).



\bibitem{Rey} S. Reynaud, B. Lamine, A. Lambrecht. P. M. Neto and M.T. Jaekel, {\em Decoherence and gravitational backgrounds}, Int. J. Mod. Phys. A 17, 1003 (2002).



\bibitem{PowPer} W. L.  Power   and   I. C. Percival, {\em Decoherence of quantum wavepackets due to interaction with conformal spacetime fluctuations}, Proc.   Roy.   Soc.   Lond.   A456, 955 (2000).







\bibitem{KoYu} P. Kok and U. Yurtsever, {\em Gravitational decoherence}, Phys.Rev. D68, 085006  (2003).

\bibitem{Breuer} H. P. Breuer,   E.   G\"okl\"u,   and  C. L\"ammerzahl, {\em Metric fluctuations and decoherence}, Class.   Quantum   Grav. 26, 105012 (2009).


\bibitem{AGB} L. Asprea, G. Gasbarri, H. Ulbricht  and A. Bassi, {\em On the decoherence effect of a stochastic gravitational perturbation on scalar matter and the possibility of its interferometric detection}, Phys. Rev. Lett. 126, 200403 (2021).


 \bibitem{Fein19} Y. Y. Fein et al, {\em Quantum Superposition of Molecules Beyond 25 kDa}, Nature Physics, 15, 1242 (2019). https://doi.org/10.1038/s41567-019-0663-9



\bibitem{Adler} S. Adler, {\em Quantum theory as an emergent phenomenon} (Cambridge University Press, Cambridge, 2004).

\bibitem{Diosi} L. Diosi, {\em Intrinsic time-uncertainties and decoherence: comparison of 4 models}, Braz. J. Phys. 35, 260  (2005).

\bibitem{LagAn} M. Lagouvardos and C. Anastopoulos, {\em Gravitational decoherence of photons}, Class. Quant. Grav. 38, 115012 (2021).


\bibitem{MAQRO}  R. Kaltenbaek et al, {\em Macroscopic Quantum Resonators (MAQRO): 2015 Update},  	EPJ Quantum Technology 3, 5 (2016).

\bibitem{MAQRO0} R. Kaltenbaek et al, {\em Macroscopic Quantum Resonators (MAQRO)}, Exp. Astron. 34, 123 (2012).



\bibitem{GiGr}  D. Giulini  and A. Großardt, {\em Gravitationally induced inhibitions of dispersion according to the Schrödinger-Newton Equation}, Class. Quantum Grav. 28, 195026 (2011).

    \bibitem{Grossa} A. Großardt, {\em Approximations for the free evolution of self-gravitating quantum particles},  	Phys. Rev. A 94, 022101 (2016).


\bibitem{DSQL} L. Mazzarella et al, {\em Goals and Feasibility of the Deep Space Quantum Link}, Proc. SPIE 11835, Quantum Communications and Quantum Imaging XIX, 118350J (2021).

    M. Mohageg, {\em The Deep Space Quantum Link: Prospective Fundamental
Physics Experiments using Long-Baseline Quantum Optics}, (ready for submission)



\bibitem{expdec}
 A. Bassi, K. Lochan, S. Satin, T. P. Singh, and H. Ulbricht, {\em Models of Wave-Function Collapse, Underlying Theories, and Experimental Tests}, Rev. Mod. Phys. 85, 471 (2013).

 G. Gasbarri, A. Belenchia, M. Carlesso et al, {\em Testing the Foundation of Quantum Physics in Space Via Interferometric and Non-Interferometric Experiments with Mesoscopic Nanoparticles}, Commun. Phys. 4, 155 (2021).


\end{thebibliography}
\end{document}